\documentclass[10pt,twocolumn,twoside]{IEEEtran}
\usepackage{graphicx}
\usepackage{amsmath}
\usepackage{array}
\newcommand{\PreserveBackslash}[1]{\let\temp=\\#1\let\\=\temp}
\newcolumntype{C}[1]{>{\PreserveBackslash\centering}p{#1}}
\newcolumntype{R}[1]{>{\PreserveBackslash\raggedleft}p{#1}}
\newcolumntype{L}[1]{>{\PreserveBackslash\raggedright}p{#1}}
\usepackage[usenames]{color}
\usepackage{colortbl,booktabs}
\usepackage{tabularx}
\usepackage{multicol}
\usepackage{booktabs}
\usepackage[Symbol]{upgreek}
\usepackage{subfigure}
\usepackage{bm}
\usepackage{cite}
\usepackage{balance}
\usepackage[boxed]{algorithm2e}
\usepackage{mathrsfs}

\usepackage{threeparttable}
\usepackage{amsmath}
\usepackage{dcolumn}
\usepackage{multirow}
\usepackage{booktabs}
\usepackage{amsfonts}
\newcolumntype{d}[1]{D{.}{.}{#1}}

\def \qed {\hfill \vrule height6pt width 6pt depth 0pt}

\begin{document}

\bibliographystyle{IEEEtran} 


\title{GMD-Based Hybrid Precoding For Millimeter-Wave Massive MIMO Systems}


\author{Tian Xie, Linglong Dai, Xinyu Gao, Muhammad Zeeshan Shakir, and Zhaocheng Wang


\thanks{T. Xie, L. Dai, X. Gao, and Zhaocheng Wang are with the Tsinghua National Laboratory
for Information Science and Technology (TNList), Department of Electronic Engineering, Beijing 100084, China (e-mail: razor9321@163.com, daill@mail.tsinghua.edu.cn, gxy1231992@sina.com, and zcwang@mail.tsinghua.edu.cn).}
\thanks{Muhammad Zeeshan Shakir is with the Department of Systems and Computer Engineering, Carleton University, Ottawa, Ontario, Canada. (e-mail: muhammad.shakir@sce.carleton.ca).}
\thanks{This work was supported in part by the International Science \& Technology Cooperation Program of China (Grant No. 2015DFG12760),
the National Natural Science Foundation of China (Grant Nos. 61571270 and 61271266),  the Beijing Natural Science Foundation (Grant No. 4142027),
and the Foundation of Shenzhen government.}}
%

\maketitle
\vspace*{-3.5mm}
\begin{abstract}
\hspace*{-1.05mm}
Hybrid precoding can significantly reduce the number of required radio frequency (RF) chains and relieve the huge energy consumption in mmWave massive MIMO systems, thus attracting much interests from academic and industry. However, most existing hybrid precoding schemes are based on singular value decomposition (SVD). Due to the very different sub-channel signal-to-noise ratios (SNRs) after SVD, complicated bit allocations is usually required to match the sub-channel SNRs. To solve this problem, we propose a geometric mean decomposition (GMD)-based hybrid precoding scheme to avoid the complicated bit allocation. Its basic idea is to seek a pair of analog and digital precoding matrices that are sufficiently close to the optimal unconstrained GMD precoding matrix. Specifically, we design the analog (digital) precoding matrix while keeping the digital (analog) precoding matrix fixed. Further, the principle of basis pursuit is utilized in the design of analog precoding matrix, while we obtain the digital precoding matrix by projecting the GMD operation on the digital precoding matrix. Simulation results verify that the proposed GMD-based hybird precoding scheme outperforms conventional SVD-based hybrid precoding schemes and achieves much better bit error rate (BER) performance with low complexity.
\end{abstract}

\vspace*{0mm}
\begin{keywords}
Massive MIMO, mmWave communications, hybrid precoding, geometric mean decomposition.
\end{keywords}

\section{Introduction}\label{S1}

Due to the vast unlicensed spectrum in millimeter-wave (mmWave) band (30-300 GHz), mmWave communications can offer huge spectrum resource \cite{bai2014coverage}. On the other hand, massive MIMO can provide enough array gain through a large number of antennas (e.g., 256 antennas) to compensate for the severe attenuation of mmWave signals \cite{rusek13}. Thus, the combination of mmWave communication and massive MIMO (mmWave massive MIMO) is a promising technology for the future 5G wireless communications, since it can significantly increase the system throughput \cite{wei2014key}. However, the conventional fully digital precoding structure, where one dedicated radio frequency (RF) chain is required for each antenna, is too energy-consumptive due to the large number of RF chains (e.g., 256 RF chains for 256 antennas) in mmWave massive MIMO systems. To address this problem, a hybrid analog/digital precoding structure has been proposed \cite{han2015large}, which decomposes the fully digital precoding matrix into one low-dimensional digital precoding matrix and one high-dimensional analog precoding realized through analog circuits (e.g., phase shifter networks) \cite{han2015large}. In this way, the hybrid precoding structure only needs a small number of RF chains to achieve the near-optimal performance \cite{han2015large}.

%
%


How to obtain the optimal precoding matrix is the key issue for hybrid precoding. In \cite{el2013spatially}, the authors formulated the hybrid precoding design problem as a sparse reconstruction problem, and proposed to use the orthogonal matching pursuit (OMP) algorithm to obtain the analog precoding matrix. \cite{rusu2015low} proposed a low-complexity version of the algorithm in \cite{el2013spatially} based on the local search method, while \cite{zhang2014achieving} derived a lower bound on the numbers of RF chains to perfectly achieve the performance of fully digital precoding by using the hybrid precoding structures. The basic ideas of these works above are seeking a pair of analog and digital precoding matrices that are sufficiently close to the right singular matrix from the singular value decomposition (SVD) of the channel matrix (which are called SVD-based hybrid precoding in this paper). SVD-based hybrid precoding can achieve the capacity-approaching performance with the help of water-filling power allocation. Nevertheless, due to the very different SNRs for each sub-channel in SVD-based precoding, complicated bit allocations (i.e., allocating different modulation and coding schemes (MCS) on different sub-channels) are needed to match the SNRs of different sub-channels, which involve high or even unaffordable coding/decoding complexity in practice \cite{chao2011bit}.


 In this paper, we propose a hybrid precoding scheme for mmWave massive MIMO systems based on the geometric mean decomposition (GMD) to avoid the complicated bit allocations. Specifically, different from the conventional SVD-based hybrid precoding, we propose to treat the right semi-unitary matrix in GMD as the optimal unconstrained precoding matrix, which can convert the mmWave massive MIMO channel into identical sub-channels and therefore naturally do not require bit allocations any more. Furthermore, to efficiently find a solution to the GMD-based hybrid precoding design problem, which is difficult to be solved, we use a decoupled optimization method to design the analog and digital precoding matrix. Particularly, we solve the analog precoding matrix design problem via the principle of basis pursuit inspired by \cite{el2013spatially}, while we obtain the digital precoding matrix by projecting the GMD operation on the digital precoding matrix. Simulation results verify that the proposed GMD-based hybird precoding scheme outperforms the conventional SVD-based hybrid precoding schemes and achieves much better bit error rate (BER) performance with low complexity.


\emph{Notation}: Lower-case and upper-case boldface letters ${\bf{a}}$ and ${\bf{A}}$ denote a vector and a matrix. ${tr \{ {\bf{A}} \} }$, ${ \{ {\bf{A}} \} }^T$, ${ \{ {\bf{A}} \} }^H$, ${ || {\bf{A}} || }_F$, and ${\{ {\bf{A}} \}}_{i,j}$  denote the trace, transpose, conjugate transpose, Frobenius norm, and the element in the \emph{i}th row and the \emph{j}th column of ${\bf{A}}$, respectively. $a := b$ is the in-place update operation, where $a$ is replaced by $b$.

\section{System Model}\label{sec:sys_model}

\subsection{Hybrid analog/digital precoding}
We consider a typical mmWave massive MIMO system with hybrid precoding structure, where the base station (BS) with $N_{\text{t}}$ transmit antennas sends $N_{\text{s}}$ independent data streams to the user with $N_{\text{r}}$ receiving antennas. Furthermore, we assume that the BS and the user have $N^{\text{RF}}_{\text{t}}$ and $N^{\text{RF}}_{\text{r}}$ RF chains, respectively, which satisfy $N_{\text{s}} \leq N^{\text{RF}}_{\text{t}} \leq N_{\text{t}}$ and $N_{\text{s}} \leq N^{\text{RF}}_{\text{r}} \leq N_{\text{r}}$ \cite{el2013spatially}. In the hybrid precoding structure, as shown in Fig. 1, the hybrid precoding matrix ${\bf{P}} \in {\mathbb{C}}^{N_{\text{t}} \times N_{\text{s}}}$ at the BS can be written as the product of two parts: the first part is a low-dimension digital precoding matrix ${\bf{P}}_{\text{D}} \in {\mathbb{C}}^{N^{\text{RF}}_{\text{t}} \times N_{\text{s}}} $; the second part is a high-dimension analog precoding matrix ${\bf{P}}_{\text{A}} \in {\mathbb{C}}^{N_{\text{t}} \times N^{\text{RF}}_{\text{t}}}$, i.e., $ {\bf{P}} = {\bf{P}}_{\text{A}}{\bf{P}}_{\text{D}}$, where ${\mathbb C}$ is the set of complex numbers. Note that the above remarks on precoding matrix can also be applied to the combining matrix. Thus, the transmitted signal vector ${\bf{x}}$ is
\begin{equation}\label{1}
{ \bf{x}} =  {\bf{P}}{\bf{s}} = {\bf{P}}_{\text{A}}{\bf{P}}_{\text{D}}{\bf{s}},
\end{equation}
where ${\bf{s}} \in {\mathbb{C}}^{N_{\text{s}} \times 1}$ denotes the source signal vector. To meet the constraint of transmitting power, we bound the transmit power at the BS as $\text{tr}\{ {\bf{P}} {\bf{P}}^H \} \leq N_{\text{s}}$. After receiving the signal vector, the user utilizes a hybrid combining matrix ${\bf{W}}$ to combine the signal vector:
\begin{equation}\label{1}
\begin{aligned}
{ \bf{y}} &=  \sqrt{\rho}{\bf{W}}^H{\bf{H}}{\bf{x}} + {\bf{W}}^H{\bf{n}} \\
          &=  \sqrt{\rho}{\bf{W}}^H_{\text{B}}{\bf{W}}^H_{\text{A}}{\bf{H}}{\bf{P}}_{\text{A}}{\bf{P}}_{\text{D}}{\bf{s}} + {\bf{W}}^H_{\text{D}}{\bf{W}}^H_{\text{A}}{\bf{n}},
\end{aligned}
\end{equation}
where $\rho$ is the average received power, ${\bf{H}} \in {\mathbb{C}}^{N_{\text{r}} \times N_{\text{t}}}$ denotes the channel matrix between BS and the user, and ${\bf{n}} \in {\mathbb{C}}^{N_{\text{r}} \times 1} $ is the additive white Gaussian noise (AWGN) vector at the user following the circularly symmetric complex Gaussian distribution with zero mean vector and covariance matrix $\sigma^2{\bf{I}}_{N_{\text{r}}}$, i.e., $\mathcal{CN}({\bf{0}},\sigma^2{\bf{I}}_{N_{\text{r}}})$, with $\sigma^2$ the noise covariance, and ${\bf{I}}_{N_{\text{r}}}$ the $N_{\text{r}}$ by $N_{\text{r}}$ identity matrix. Since the analog precoder is realized through analog phase shifter network after up-converters, all elements of ${\bf{P}}_{\text{A}}$ should have the same amplitude:
\begin{equation}
 |  \{{\bf{P}}_{\text{A}}\}_{i,j} |  = \frac{1}{\sqrt{N_{\text{t}}}},
\end{equation}
where $| \cdot |$ denotes the modulus of a complex number.


\begin{figure}[tp]
\begin{center}
\vspace*{0mm}\includegraphics[width = 1\linewidth]{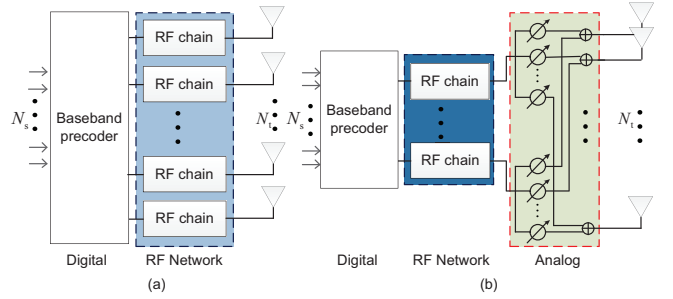}
\end{center}
\vspace*{-4mm}\caption{Comparison between two typical precoding structures: (a) The fully digital precoding; (b) The hybrid analog/digital precoding.} \label{fig1}
\vspace*{0mm}
\end{figure}

\vspace{-3mm}

\subsection{Channel Model}
To capture the limited spatial selectivity or scattering characteristics in mmWave massive MIMO channel, we adopt a widely used Saleh-Valenzuela (SV) model in this paper \cite{el2013spatially,gao2016energy}, where the channel matrix ${\bf{H}}$ can be expressed as:
\begin{equation}\label{sv}
{\bf{H}} = \sqrt {\frac{{{N_{\text{t}}}{N_{\text{r}}}}}{L}} \sum\limits_i^L {{\beta _i}{{\bf{a}}_{\text{r}}}(\varphi _i^{\text{r}}){{\bf{a}}_{\text{t}}}(\varphi _i^{\text{t}})},
\end{equation}
where $L$ denotes the number of paths, $\beta_{i}$ is the complex gain of the $i$th path, ${{\bf{a}}_{\text{r}}}(\varphi _i^{\text{r}})$ and ${{\bf{a}}_{\text{t}}}(\varphi _i^{\text{t}})$ are the array response vectors at the user and the BS, respectively, where $\varphi _i^{\text{r}}$ denotes the angle of arrival (AoA) at the user, and $\varphi _i^{\text{t}}$ is the angle of departure (AoD) at the BS. For the simple uniform linear line (ULA) antenna array of $N$ elements, the array response vector is
\begin{equation}
{{\bf{a}}_{\text{ULA}}}(\varphi ) = \sqrt {\frac{1}{N}} {\left[ 1,{e^{jkd\sin (\varphi )}},\cdots,{e^{j(N - 1)kd\sin (\varphi )}} \right] ^T},
\end{equation}
where $k = \frac{2\pi}{\lambda}$ with ${\lambda}$ the wavelength, and $d$ denotes the antenna spacing. Because of the limited spatial scattering in mmWave propagation, the mmWave massive MIMO channel ${\bf{H}}$ is low-rank \cite{gao2015mmwave}. As a result, we can leverage a finite number of RF chains to achieve the near-optimal throughput \cite{gao2015mmwave}.
%

\section{Proposed GMD-Based Hybrid Precoding}

In this section, we first review the fully digital precoding schemes. Next, we propose the GMD-based hybrid precoding.

\vspace{0mm}
\subsection{Fully digital SVD-and GMD-based precoding}
In this subsection, we first briefly review the fully digital precoding schemes in mmWave massive MIMO systems. Denote the SVD of the channel matrix ${\bf{H}}$ by
\begin{equation}
\begin{aligned}
{\bf{H}} &= {\bf{U} \Sigma V}^H \\
         &= \left[ {\begin{array}{*{20}{c}}
             {{{\bf{U}}_1}}&{{{\bf{U}}_2}}
             \end{array}} \right]\left[ {\begin{array}{*{20}{c}}
             {{{\bf{\Sigma }}_1}}&{\bf{0}}\\
             {\bf{0}}&{{{\bf{\Sigma }}_2}}
             \end{array}} \right]\left[ {\begin{array}{*{20}{c}}
             {{\bf{V}}_1^H}\\
             {{\bf{V}}_2^H}
             \end{array}} \right] \\
         &= {{\bf{U}}_1}{{\bf{\Sigma }}_1}{\bf{V}}_1^H + {{\bf{U}}_2}{{\bf{\Sigma }}_2}{\bf{V}}_2^H,
\end{aligned}
\end{equation}
where ${{{\bf{U}}_1}} \in {\mathbb{C}}^{N_{\text{r}} \times N_{\text{s}}} $ and ${{{\bf{V}}_1}} \in {\mathbb{C}}^{N_{\text{t}} \times N_{\text{s}}} $ are semi-unitary matrices containing the left $N_{\text{s}}$ columns of unitary matrices ${\bf{U}} \in {\mathbb{C}}^{N_{\text{r}} \times N_{\text{r}}}$ and ${\bf{V}} \in {\mathbb{C}}^{N_{\text{t}} \times N_{\text{t}}}$, respectively, and ${{\bf{\Sigma }}_1} = \text{diag}({\sigma _1}, \cdots ,{\sigma _{{N_{\text{s}}}}})$ is a diagonal matrix with the largest $N_{\text{s}}$ singular values of ${\bf{H}}$ in decreasing order. With fully digital SVD-based precoding, i.e., ${\bf{P}} = {\bf{V}}_{1}$ and $ {\bf{W}} = {\bf{U}}^{H}_{1}$, the MIMO channels are converted into $N_{\text{s}}$ parallel sub-channels, of which the sub-channel gains are ${\sigma _1}, \cdots ,{\sigma _{{N_{\text{s}}}}}$, i.e.,
\begin{equation}\label{svd_2}
\begin{aligned}
{ \bf{y}}  &= \sqrt{\rho}{\bf{W}}^H{\bf{H}}{\bf{P}}{\bf{s}} + {\bf{W}}^H{\bf{n}} \\
           &= \sqrt{\rho}{\bf{U}}_{1}^{H}{\bf{H}}{\bf{V}}_{1}{\bf{s}} + {\bf{U}}_{1}^{H}{\bf{n}}=  \sqrt{\rho}{\bf{\Sigma}}_{1} {\bf{s}} + {\bf{U}}_{1}^{H}{\bf{n}}.
\end{aligned}
\end{equation}
Due to the limited spatial scattering in mmWave propagation, the singular values of channel matrix ${\bf{H}}$ varies a lot \cite{gao2015mmwave}, which results in the very different SNRs among different sub-channels after water-filling power allocation\footnote{Note that the water-filling algorithm allocates less power on the sub-channel with lower channel gain, which worsens the SNR for this sub-channel.} as shown in Fig. 2 (a). Consequently, if we use the same MCS for all sub-channels, the BER performance for the MIMO system will be determined by the sub-channel with the lowest SNR. On the other hand, if we want to guarantee the similar BER performance among all sub-channels, careful bit allocations are needed, which can bring the unaffordable coding/decoding complexity in mmWave massive MIMO systems \cite{chao2011bit}.

To solve this problem, GMD has been proposed \cite{jiang2005geometric}, where the channel matrix ${\bf{H}}$ is decomposed as:
\begin{equation}
\begin{aligned}
{\bf{H}} = {\bf{GRQ}}^H = \left[ {\begin{array}{*{20}{c}}
        {{{\bf{G}}_1}}&{{{\bf{G}}_2}}
        \end{array}} \right]\left[ {\begin{array}{*{20}{c}}
         {{{\bf{R}}_1}}&{\bf{*}}\\
        {\bf{0}}&{{{\bf{R}}_2}}
         \end{array}} \right]\left[ {\begin{array}{*{20}{c}}
         {{\bf{Q}}_1^H}\\{{\bf{Q}}_2^H} \end{array}} \right],
\end{aligned}
\end{equation}
where ${{{\bf{G}}_1}} \in {\mathbb{C}}^{N_{r} \times N_{s}} $ and ${{{\bf{Q}}_1}} \in {\mathbb{C}}^{N_{t} \times N_{s}} $ are semi-unitary matrices containing the left $N_s$ columns of unitary matrices ${{\bf{G}}} \in {\mathbb{C}}^{N_{r} \times N_{r}} $ and ${{\bf{Q}}} \in {\mathbb{C}}^{N_{t} \times N_{t}} $, ${\bf{R}}_1$ is an upper triangular matrix with identical diagonal elements\footnote{Here, we use the simple form $r_{i,j}$ to denote the element in the \emph{i}th row and \emph{j}th row of ${\bf{R}}_1$ for simplicity.}, i.e., the geometric mean of singular values ${r_{i,i}} = ( \sigma_1 \sigma_2 \cdots \sigma_{N_{\text{s}}}  )^{\frac{1}{{{N_{\text{s}}}}}} = {\bar r}, \forall i$, and ${\bf{*}}$ denotes an arbitrary matrix that we do not care. Note that the diagonal element in ${\bf{R}}_1$ is the geometric mean of the first $N_{\text{s}}$ singular values of ${\bf{H}}$, since there are only $N_s$ data streams. The fully digital GMD-based precoding employs ${\bf{Q}}_1$ as the precoding matrix and ${\bf{G}}_1^H$ as the combining matrix, where we can transform the effective channel matrix in (2) into an upper triangular matrix:
\begin{equation}
\begin{aligned}
{ \bf{\hat x}}  &= {\bf{W}}^H{\bf{H}}{\bf{P}}{\bf{s}} + {\bf{W}}^H{\bf{n}} \\
                &= {\bf{G}}_{1}^{H}{\bf{H}}{\bf{Q}}_{1}{\bf{s}} + {\bf{G}}_{1}^{H}{\bf{n}}=  {\bf{R}}_{1} {\bf{s}} +  {\bf{G}}_{1}^{H}{\bf{n}}.
\end{aligned}
\end{equation}
Then, by utilizing successive interference cancellation (SIC) at the receiver \cite{jiang2005geometric}, we can acquire $N_s$ sub-channels with equal sub-channel gain $r_{ii}$ as shown in Fig. 2 (b), which means that we can avoid the bit allocation caused by different SNRs for different sub-channels in precoding schemes based on SVD.

\begin{figure}[tp]
\begin{center}
\vspace*{0mm}\includegraphics[width = 0.95\linewidth]{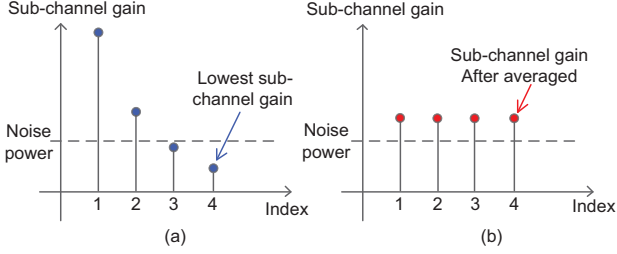}
\end{center}
\vspace*{-4mm}\caption{Intuitive illustration of sub-channel gains: (a) Conventional SVD-based hybrid precoding; (b) Proposed GMD-based hybrid precoding.} \label{fig1}
\vspace*{0mm}
\end{figure}

\vspace{-3mm}

\subsection{GMD-based hybrid precoding}
In this subsection, we turn to discuss the GMD-based hybrid precoding. According to (1) and (4), we can formulate the hybrid precoding problem as follow\footnote{To distinguish the GMD-based hybrid precoding from the SVD-based precoding, we use ${\bf{Q}}_1$,  ${\bf{Q}}_{\text{A}}$, and ${\bf{Q}}_{\text{D}}$ to denote the unconstrained GMD-based precoding matrix, the GMD-based analog precoding matrix, and the GMD-based digital precoding matrix, respectively.} \cite{el2013spatially}:
\begin{equation}\label{11}
\begin{array}{l}
\begin{aligned}
\left( {\bf{Q}}_{\text{A}}^{\text{opt}},{\bf{Q}}_{\text{D}}^{\text{opt}} \right) &= \mathop {\arg \min }\limits_{{\bf{Q}}_{\text{A}},{\bf{Q}}_{\text{D}}} ||{{\bf{Q}}_1} - {\bf{Q}}_{\text{A}}{\bf{Q}}_{\text{D}}|{|_F},\\
&\text{s.t.} \ \ |\{{\bf{Q}}_{\text{A}}\}_{i,j}| = \frac{1}{\sqrt{N_{\text{t}}}},\\
& \ \ \ \ \ tr \left( {\bf{Q}}_{\text{A}}{\bf{Q}}_{\text{D}} {\bf{Q}}^H_{\text{D}}{\bf{Q}}^H_{\text{A}} \right) \le {N_{\text{s}}}.
\end{aligned}
\end{array}
\end{equation}
This problem formulation can be intuitively understood as seeking a pair of ${\bf{Q}}_{\text{A}}{\bf{Q}}_{\text{D}}$ sufficiently ``close'' to the optimal unconstrained precoding matrix ${\bf{Q}}_{1}$.

%


The following \textbf{Lemma 1} shows some insights about how to efficiently solve (\ref{11}).

\vspace*{+2mm} \noindent\textbf{Lemma 1}. {\it The GMD of ${\bf{H}}$ can be expressed as a rotation of SVD as following
\begin{equation}
\begin{aligned}
{\bf{Q}}_1 = {\bf{V}}_1 {\bf{S}}_R;\  {\bf{G}}_1 = {\bf{U}}_1 {\bf{S}}_L;\ {\bf{R}}_1 = {\bf{S}}^T_L{\bf{\Sigma}}_1 {\bf{S}}_R
\end{aligned}
\end{equation}
where ${\bf{S}}_R \in {\mathbb{C}}^{N_{s} \times N_{s}} $ and ${\bf{S}}_L \in {\mathbb{C}}^{N_{s} \times N_{s}} $ are unitary matrices dependent on ${\bf{\Sigma}}_1$.}

\vspace*{+2mm}

\textit{Proof:} According to \cite{jiang2005joint}, the implementation algorithm of GMD based on SVD can be briefly summarized as follow:
\subsubsection{Step1} Initialize ${\bf{R}}_1 = {\bf{\Sigma}}_1$, ${\bf{Q}}_1 = {\bf{V}}_1$, ${\bf{G}}_1 = {\bf{U}}_1$, and compute the geometric mean value ${\bar r} = ( \sigma_1 \sigma_2 \cdots \sigma_{N_{\text{s}}}  )^{\frac{1}{{{N_{\text{s}}}}}}$.
\subsubsection{Step2} At stage $i$, where $i$ varies from $1$ to $N_s - 1$, first check the \emph{i}th diagonal element of ${\bf{R}}_1$, i.e., $r_{i,i}$: If $r_{i,i} \geq \bar r$, find certain element $r_{p,p}$ such that $r_{p,p} \leq \bar r$, where $p > i$; Otherwise, find certain element $r_{p,p}$ such that $r_{p,p} \geq \bar r$, where $p > i$. Then, swap $r_{i+1,i+1}$ and $r_{p,p}$, as well as the corresponding (\emph{i+1})th and \emph{p}th column in ${\bf{Q}}_1$ and ${\bf{G}}_1$ by
\begin{equation}
\begin{aligned}
{\bf{R}}_1:=  \{ {\bf{M}}^{(i)} \}^{T} {\bf{R}}_1  {\bf{M}}^{(i)}, \ {\bf{Q}}_1:=  {\bf{Q}}_1 {\bf{M}}^{(i)}, \ {\bf{G}}_1 :=  {\bf{G}}_1 {\bf{M}}^{(i)},
\end{aligned}
\end{equation}
where ${\bf{M}}^{(i)}$ is the corresponding permutation matrix.

\subsubsection{Step3} Construct two Givens matrices ${\bf{N}}^{(i)}_{\text{L}}$ and ${\bf{N}}^{(i)}_{\text{R}}$ from ${\bf{I}}_{N_{\text{s}}}$ by replacing the sub-matrix containing the four elements $\{{\bf{N}}^{(i)}_{\text{L}}\}_{i,i}$, $\{{\bf{N}}^{(i)}_{\text{L}}\}_{i+1,i}$, $\{{\bf{N}}^{(i)}_{\text{L}}\}_{i,i+1}$, and $\{{\bf{N}}^{(i)}_{\text{L}}\}_{i+1,i+1}$ with the following $2 \times 2$ matrices:
\begin{equation}
\begin{aligned}
{\bf{\Theta}}^{(i)}_{\text{L}} =  \frac{1}{\bar r }\left[\begin{array}{*{10}{c}} {c{r_{i,i}}}&{s{r_{i + 1,i + 1}}}\\ { - s{r_{i + 1,i + 1}}}&{c{r_{i,i}}} \end{array}\right],
{\bf{\Theta}}^{(i)}_{\text{R}} =  \left[\begin{array}{*{20}{c}} {c}&{-s}\\ { s}&{c} \end{array}\right]
\end{aligned}
\end{equation}
respectively, where
\begin{equation}
c = \sqrt{\frac{ {\bar r}^2 - r^2_{i+1,i+1} }{r^2_{i,i} - r^2_{i+1,i+1}}}, \ s = \sqrt{1 - c^2}.
\end{equation}
Then, update ${\bf{R}}_1$, ${\bf{Q}}_1$, and ${\bf{G}}_1$ as
\begin{equation}\label{update}
\begin{aligned}
{\bf{R}}_1:=   {\bf{N}}^{(i)}_{\text{L}} {\bf{R}}_1 {\bf{N}}^{(i)}_{\text{R}},
{\bf{Q}}_1:=  {\bf{Q}}_1  \{{\bf{N}}^{(i)}_{\text{L}}\}^T,
{\bf{G}}_1:=  {\bf{G}}_1  {\bf{N}}^{(i)}_{\text{R}}.
\end{aligned}
\end{equation}
One can verify that
\begin{equation}
{\bf{\Theta}}^{(i)}_{\text{L}} \left[\begin{array}{*{20}{c}} {r_{i,i}}&{0}\\ {0}&{r_{i+1,i+1}} \end{array}\right] {\bf{\Theta}}^{(i)}_{\text{R}} = \left[\begin{array}{*{20}{c}} {\bar r}&{*}\\ {0}&{\frac{r_{i,i}r_{i+1,i+1}}{\bar r}} \end{array}\right],
\end{equation}
which means that by ${\bf{N}}^{(i)}_{\text{L}}{\bf{R}}_1 {\bf{N}}^{(i)}_{\text{R}}$, we can make $r_{i,i} = \bar r$, while keeping other elements unchanged.

\subsubsection{Step4} Update $i := i+1$ and go back to Step 2 until $i$ equals $N_s - 1$.

Combining (\ref{update}) and the fact that ${\bf{M}}^{(i)}$, ${\bf{N}}^{(i)}_{\text{L}}$, and ${\bf{N}}^{(i)}_{\text{R}}$ are all unitary matrices, we can confirm \textbf{Lemma 1} by letting ${\bf{S}}_{\text{L}} = {\bf{M}}^{(1)} {\bf{N}}^{(1)}_{\text{R}} {\bf{M}}^{(2)} {\bf{N}}^{(2)}_{\text{R}} \cdots {\bf{M}}^{(N_{\text{s}})} {\bf{N}}^{(N_{\text{s}})}_{\text{R}}$ and ${\bf{S}}_{\text{R}} = {\bf{M}}^{(1)} \{{\bf{N}}^{(1)}_{\text{L}}\}^T {\bf{M}}^{(2)} \{{\bf{N}}^{(2)}_{\text{L}}\}^T \cdots {\bf{M}}^{(N_{\text{s}})} \{{\bf{N}}^{(N_{\text{s}})}_{\text{L}}\}^T$. \qed

According to \textbf{Lemma 1}, the objective function in (\ref{11}) can be written as
\begin{equation}\label{18}
\begin{aligned}
||{{\bf{Q}}_1} - {\bf{Q}}_{{\text{A}}}{\bf{Q}}_{{\text{D}}}|{|_F} &= || {\bf{V}}_1 {\bf{S}}_{{\text{R}}} - {\bf{Q}}_{{\text{A}}}{\bf{Q}}_{{\text{B}}} ||_F \\
                                                            &\overset{(a)}{=} || {\bf{V}}_1  - {\bf{Q}}_{{\text{A}}} {\bf{Q}}_{{\text{D}}} \{ {\bf{S}}_{\text{R}}\} ^{H}  ||_F \\
                                                            &= || {\bf{V}}_1  - {\bf{Q}}_{{\text{A}}} \widetilde {\bf{Q}}_{{\text{D}}}  ||_F,
\end{aligned}
\end{equation}
where (a) holds because the Frobenius norm is invariant under rotations, and $\widetilde {\bf{Q}}_{\text{D}} = {\bf{Q}}_{\text{D}} \{ {\bf{S}}_{\text{R}}\} ^{H}$. From (\ref{18}), we can observe that the optimal unconstrained precoding matrix in (\ref{11}) will be ${\bf{V}}_1$, if we rotate ${\bf{Q}}_{\text{D}}$ as $\widetilde {\bf{Q}}_{\text{D}} = {\bf{Q}}_{\text{D}} \{ {\bf{S}}_{\text{R}}\} ^{H}$. So we reformulate (\ref{11}) as
\begin{equation}\label{20}
\begin{array}{l}
\begin{aligned}
\left( {\bf{Q}}_{\text{A}}^{\text{opt}},{\bf{Q}}_{\text{D}}^{\text{opt}} \right) &= \mathop {\arg \min }\limits_{{\bf{Q}}_{\text{A}}, \widetilde {\bf{Q}}_{\text{D}}}    || {\bf{V}}_1  - {\bf{Q}}_{{\text{A}}} \widetilde {\bf{Q}}_{{\text{D}}}  ||_F,\\
&\text{s.t.} \ \ |\{{\bf{Q}}_{\text{A}}\}_{i,j}| = \frac{1}{\sqrt{N_{\text{t}}}},\\
& \ \ \ \ \ tr \left( {\bf{Q}}_{\text{A}} \widetilde {\bf{Q}}_{\text{D}} {\bf{Q}}^H_{\text{D}} \widetilde {\bf{Q}}^H_{\text{A}}  \right) \le {N_{\text{s}}}.
\end{aligned}
\end{array}
\end{equation}


However, solving (\ref{20}) is also highly complicated, since ${\bf{Q}}_{{\text{A}}}$ and $\widetilde {\bf{Q}}_{{\text{D}}}$ are coupled, and the elemental-wise constraints $|\{{\bf{Q}}_{\text{A}}\}_{i,j}| = {1}/{\sqrt{N_{\text{t}}}}$ is non-convex \cite{el2013spatially}. To this end, we propose to solve to (\ref{20}) by decoupling the design ${\bf{Q}}_{\text{A}}$ and ${\bf{Q}}_{\text{D}}$, i.e., treating ${\bf{Q}}_{\text{D}}$ as a fixed matrix while designing ${\bf{Q}}_{\text{A}}$, and vice versa. Note that the main difficult lies in the design of the analog precoding matrix ${\bf{Q}}_{\text{A}}$ due to a non-convex constraint on ${\bf{Q}}_{\text{A}}$. To effectively design the analog precoding matrix, inspired by \cite{el2013spatially}, we leverage the array response vectors ${{\bf{a}}_{\text{t}}}(\varphi _i^{\text{t}}), \forall i$ as the columns of ${\bf{Q}}_{\text{A}}$, which can be explained as follow. Firstly, ${{\bf{a}}_{\text{t}}}(\varphi _i^{\text{t}}), \forall i$ has constant modulus elements, which satisfy the constraint $|\{{\bf{Q}}_{\text{A}}\}_{i,j}| = {1}/{\sqrt{N_{\text{t}}}}$. Secondly, the columns of ${\bf{V}}_1$ form an orthogonal basis of the channel's row space, while from (\ref{sv}), the array response vectors ${{\bf{a}}_{\text{t}}}(\varphi _i^{\text{t}}), \forall i$ also form an orthogonal basis of the channel's row space. As a result, ${\bf{V}}_1$ can be expressed as a linear combination of ${{\bf{a}}_{\text{t}}}(\varphi _i^{\text{t}}), \forall i$. In addition, the hybrid precoding matrix ${\bf{Q}}_{\text{A}} \widetilde {\bf{Q}}_{\text{D}}$ can be also seen as a linear combination of ${\bf{Q}}_{\text{A}}$. Since we want to approximate ${\bf{V}}_1$ via ${\bf{Q}}_{\text{A}} \widetilde {\bf{Q}}_{\text{D}}$, it is reasonable to find the ``best'' $N_{\text{t}}^{\text{RF}}$ array response vectors from ${{\bf{a}}_{\text{t}}}(\varphi _i^{\text{t}}), \forall i$ as the columns of ${\bf{Q}}_{\text{A}}$. So the corresponding analog precoding design problem becomes
\begin{equation}\label{22}
\begin{array}{l}
\begin{aligned}
 {\bf{T}}  &= \mathop {\arg \min }\limits_{{\bf{T}}}    || {\bf{V}}_1  -  {\bf{A}}_{\text{t}} {\bf{T}} \widetilde {\bf{Q}}_{{\text{D}}}  ||_F,\\
&\text{s.t.} \ \    ||\text{diag}({\bf{T}} {\bf{T}}^H)||_{0} = N_{\text{t}}^{\text{RF}},\\
& \ \ \ \ \ \text{tr} \left( {\bf{A}}_{\text{t}} {\bf{T}} \widetilde {\bf{Q}}_{{\text{D}}} \widetilde {\bf{Q}}^H_{{\text{D}}} {\bf{T}}^H {\bf{A}}^H_{\text{t}} \right) \le {N_{\text{s}}},
\end{aligned}
\end{array}
\end{equation}
where ${\bf{A}}_{\text{t}} = \left[ {\bf{a}}_t(\phi^t_1), {\bf{a}}_t(\phi^t_2), \cdots, {\bf{a}}_t(\phi^t_L)  \right]$ is an $N_t \times L$ matrix, and ${\bf{T}}$ is a selecting matrix with $N_{\text{t}}^{\text{RF}}$ non-zero rows. Problem (\ref{22}) is a sparse reconstruction problem, which can be effectively solved via the principle of basis pursuit \cite{el2013spatially}. After we determine the analog precoding matrix, the digital precoding matrix design problem is a Frobenius norm minimization problem:
\begin{equation}\label{23}
\begin{array}{l}
\begin{aligned}
{\bf{Q}}_{\text{D}}^{\text{opt}}  &= \mathop {\arg \min }\limits_{ \widetilde {\bf{Q}}_{\text{D}}}    || {\bf{V}}_1  - {\bf{Q}}_{{\text{A}}} \widetilde {\bf{Q}}_{{\text{D}}}  ||_F,\\
&\text{s.t.} \ \
\text{tr} \left( {\bf{Q}}_{\text{A}} \widetilde {\bf{Q}}_{\text{D}} {\bf{Q}}^H_{\text{D}} \widetilde {\bf{Q}}^H_{\text{A}}  \right) \le {N_{\text{s}}}.
\end{aligned}
\end{array}
\end{equation}
The optimal solution to (\ref{23}) has a least square form \cite{gao2016energy}
\begin{equation}\label{baseband}
 \widetilde {\bf{Q}}_{\text{D}} =  \{{\bf{Q}}_{\text{A}}\}^{\dagger}  {\bf{V}}_1.
\end{equation}
Furthermore, we have
\begin{equation}\label{baseband}
{\bf{Q}}_{\text{D}} =  \widetilde {\bf{Q}}_{\text{D}} {\bf{S}}_{\text{R}} =  \{{\bf{Q}}_{\text{A}}\}^{\dagger}  {\bf{V}}_1  {\bf{S}}_{\text{R}}.
\end{equation}
which can be seen as projecting the GMD operation on the digital precoding matrix $\widetilde {\bf{Q}}_{\text{D}}$ to (\ref{23}).


\begin{algorithm}[tp]
\caption{The proposed GMD-based hybrid precoding}
\textbf{Require:} {\bf{H}};
    \\1) Perform SVD of channel: [${\bf{U}}_1$ ${\bf{\Sigma}}_1$ ${\bf{V}}_1$] = SVD(${\bf{H}}$);
    \\2) Initialize ${\bf{Q}}_{\text{A}}$ = Empty matrix, and ${\bf{Q}}_{\text{res}} = {\bf{V}}_1$
\\\textbf{For} ${i \leq N_{t}^{RF}}$
   \\ \hspace{+5mm} 3) ${\bf{\Phi}} = {\bf{A}}_{\text{t}}^H {\bf{Q}}_{\text{res}}$ ;
   \\ \hspace{+5mm} 4) $k = \text{argmax}_{l = 1,\cdots, L} ({\bf{\Phi}}{\bf{\Phi}}^H)_{l,l}$;
   \\ \hspace{+5mm} 5) ${\bf{Q}}_{\text{A}} = [{\bf{Q}}_{\text{A}} | {\bf{A}}_t^{(k)}]$;
   \\ \hspace{+5mm} 6) ${\bf{Q}}_{\text{D}} = {\bf{Q}}_{\text{A}}^{\dagger}  {\bf{V}}_1$;
   \\ \hspace{+5mm} 7) ${\bf{Q}}_{\text{res}} =  \frac{{\bf{V}}_1 -{\bf{Q}}_{\text{A}} {\bf{Q}}_{\text{D}}} {||{\bf{V}}_1 -{\bf{Q}}_{\text{A}} {\bf{Q}}_{\text{D}}||_F}$;
\\\textbf{End for}
\\    8) Generate ${\bf{S}}_{\text{R}}$ based on the algorithm in \textbf{Lemma} 1, \\ \hspace{+3mm} and ${\bf{Q}}_{\text{D}} = {\bf{Q}}_{\text{D}} {\bf{S}}_{\text{R}}$;
\\    9) Normalize ${\bf{Q}}_{\text{D}} = {\sqrt{N_{\text{s}}}} \frac{{\bf{Q}}_{\text{D}} } { || {\bf{Q}}_{\text{A}} {\bf{Q}}_{\text{D}} ||_{F}}$;
\\ 10) Return ${\bf{Q}}_{\text{A}}$ and ${\bf{Q}}_{\text{D}}$;
\end{algorithm}

The overall algorithm for GMD-based hybrid precoding is summarized in \textbf{Algorithm 1}. \textbf{Algorithm 1} can be divided into two parts in general. The first part includes step 1 to step 7, which performs conventional spatially sparse hybrid precoding \cite{el2013spatially} based on the optimal unconstrained precoding matrix ${\bf{V}}_1$. The second part includes step 8 and step 9, which performs the GMD operation on the digital precoding matrix ${\bf{Q}}_{\text{D}}$ and normalizes the effective precoding matrix to meet the transmitting power constraints. Note that in step 8, we do not need to explicitly compute ${\bf{S}}_{\text{R}}$, instead, we just apply the corresponding permutation and multiplication operations in each stage on ${\bf{Q}}_{\text{D}}$. Note that the complexity of implementing GMD given SVD, i.e., Step 8, is only $\mathcal {O} ( ({N}_{\text{s}}+{N}_{\text{t}}) {N}_{\text{s}})$ \cite{jiang2005joint}, since only a $N_{s}$ by $N_{s}$ rotation matrix ${\bf{S}}_{\text{R}}$ is multiplied to ${\bf{Q}}_{\text{D}}$, while the complexity of hybrid precoding algorithm, i.e., Step 1-7, is $\mathcal {O} ({N}^2_{\text{RF}}{N}_{\text{t}}{N}_{\text{s}})$, since we need to compute the pseudo-inverse matrix of ${\bf{Q}}_{\text{A}}$, which indicates that the GMD-based hybrid precoding will only incur small extra complexity compared with conventional SVD-based hybrid precoding.


\section{Simulation Results}
In this section, we will evaluate the performance of the proposed GMD-based hybrid for mmWave massive MIMO communications. Consider typical mmWave massive MIMO systems where $N_{\text{t}} = 128$-and $N_{\text{t}} = 256$-element ULAs with the antenna spacing $d = \lambda/2$ are utilized at the BS, while $N_{\text{r}}=16$-element ULA with also the antenna spacing $d = \lambda/2$ antennas is utilized at the user. The BS and the user both adopt $N_{\text{t}}^{\text{RF}} = N_{\text{r}}^{\text{RF}} = 4$ RF chains, and the carrier frequency is 28GHz. For the channel model, we utilize the SV model, where the number of path is set to $L = 4$, the complex gain of each path follows the distribution $\mathcal{CN}(0,1)$, and the AoAs and AoDs are uniformly distributed in $\left[ -\pi/2, \pi/2 \right]$. Finally, we use the water-filling power allocation scheme at the BS \cite{el2013spatially}, while the modulation scheme is 16QAM, which is utilized in all sub-channels for both SVD and GMD-based precoding schemes.

\begin{figure}[tp]
\begin{center}
\vspace*{-3mm}\includegraphics[width = 0.95\linewidth]{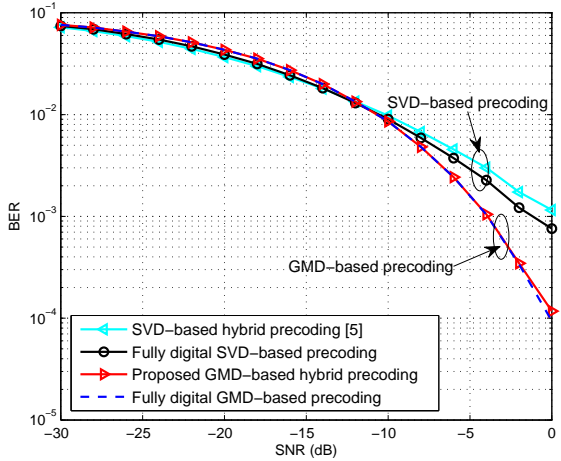}
\end{center}
\vspace*{-4mm}\caption{BER performance comparison of the proposed GMD-based hybrid precoding in a $128 \times 16$ mmWave massive MIMO system.} \label{fig3}
\vspace*{0mm}
\end{figure}

\begin{figure}[tp]
\begin{center}
\vspace*{-3mm}\includegraphics[width = 0.95\linewidth]{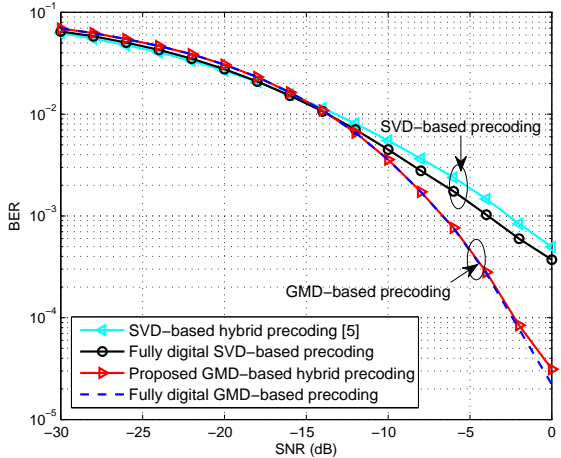}
\end{center}
\vspace*{-4mm}\caption{BER performance comparison of the proposed GMD-based hybrid precoding in a $256 \times 16$ mmWave massive MIMO system.} \label{fig4}
\vspace*{0mm}
\end{figure}

Fig. \ref{fig3} shows the BER performance of the proposed GMD-based hybrid precoding in a $128 \times 16$ mmWave massive MIMO system, where we can observe that the proposed GMD-based precoding schemes (including fully digital GMD-based precoding and GMD-based hybrid precoding) can achieve a better BER performance than the conventional SVD-based precoding (including fully digital SVD-based precoding and SVD-based hybrid precoding \cite{el2013spatially}). This is because the GMD-based precoding  converts the mmWave MIMO channel into multiple sub-channels with the identical SNR, which naturally avoids the severe BER performance degradation over sub-channels with very lowest SNR for conventional SVD-based precoding. Specifically, given the same BER performance, e.g., $\text{BER} = 10^{-3}$, the required SNR for GMD-based precoding schemes is about $-4$dB, which is lower than that of conventional SVD-based precoding. Furthermore, we can also see that the performance gap between the fully digital GMD-based precoding and the proposed GMD-based hybrid precoding is negligible, which implies that the proposed GMD-based hybrid precoding is able to well approximate the fully digital GMD-based precoding.


Fig. \ref{fig4} is the BER performance of the proposed GMD-based hybrid precoding in a $256 \times 16$ mmWave massive MIMO system. In Fig. \ref{fig4}, the BER of the proposed GMD-based precoding and the conventional SVD-based precoding both improves, since more antennas at the BS provide more array gains. However, the proposed GMD-based precoding still outperforms the conventional SVD-based precoding. The required SNR in GMD-based hybrid precoding to achieve $\text{BER} = 10^{-3}$ is about $2.5$dB and $3.5$dB lower than that in the fully digital SVD-based precoding and the SVD-based hybrid precoding \cite{el2013spatially}. In addition, the proposed GMD-based hybrid precoding can still perform close to the the fully digital GMD-based precoding with negligible gap.


\section{Conclusions}
In this paper, we propose a GMD-based hybrid precoding scheme for mmWave massive MIMO systems to avoid the complicated bit allocations in the conventional SVD-based hybrid precoding. With the help of SIC at the receiver, we can convert the mmWave MIMO channel into multiple identical sub-channels, which naturally do not need bit allocations. Specifically, we decouple the design of the analog and digital precoding matrix. The analog precoding matrix is designed using the principle of basis pursuit, while we project the GMD operation on the digital precoding matrix to obtain the digital precoding matrix. Simulation results verify that the proposed GMD-based hybird precoding scheme outperforms conventional SVD-based hybrid precoding schemes and is able to achieve much better BER performance.

\end{document}